\documentclass[10pt]{article} 

\usepackage[preprint]{rlj} 

\usepackage{amssymb}            
\usepackage{mathtools}          
\usepackage{mathrsfs}           
\usepackage{graphicx}           
\usepackage{subcaption}         
\usepackage[space]{grffile}     
\usepackage{url}                
\usepackage{lipsum}             

\usepackage{amsmath}
\usepackage{float}  
\usepackage{bm}  

\usepackage{caption} 
\usepackage{booktabs}

\title{An Agent-Centric Dynamical Systems Perspective on Multi-Agent Reinforcement Learning}

\setrunningtitle{An Agent-Centric Dynamical Systems Perspective on MARL}

\author{James Rudd-Jones\textsuperscript{1}, María Pérez-Ortiz\textsuperscript{1}, Mirco Musolesi\textsuperscript{1,2}}

\emails{\{james.rudd-jones.22,maria.perez,m.musolesi\}@ucl.ac.uk}

\affiliations{
$^{1}$\textbf{Centre for Artificial Intelligence, Department of Computer Science, University College London, London, UK}\\
$^{2}$\textbf{Department of Computer Science and Engineering, University of Bologna, Bologna, Italy}\\
}

\contribution{We formalise individual MARL learning updates as coupled discrete time dynamical systems in parameter space, covering both deterministic and stochastic settings.}
{
Prior dynamical analyses of MARL typically focus on population level approximations such as replicator dynamics \citep{maynard1982evolution, hofbauer1998evolutionary, tuyls2006evolutionary}, which average out stochasticity and algorithmic structure. 
In contrast, our formulation models the agent-level parameter updates of individual agents and their interactions with the environment as a coupled stochastic dynamical system, enabling analysis of practical MARL algorithms including stochasticity.
}
\contribution{We introduce a practical methodology for diagnosing stability and sensitivity in deep MARL using tools from dynamical systems theory, even when closed form analysis is impossible, enabling a comprehensive characterisation of the system’s dynamical regimes, transitions, and stability properties.}
{
We use classical dynamical systems tools to provide insight into system stability and attractor structure \citep{farmer1982information, lyapunov1992general, eckmann1995recurrence, grassberger1984dimensions}, enabling practitioners to characterise fixed points, cycles, and chaotic regimes.
}
\contribution{We empirically demonstrate the approach across multiple MARL games and learning algorithms, analysing regime transitions and hyperparameter sensitivity.}
{
Our work provides a principled pragmatic empirical approach to identify attractor structures, oscillatory behaviour, and transitions between dynamical regimes under different learning rules and hyperparameters.
}


\summary{
Analysing learning in Multi-Agent Reinforcement Learning (MARL) environments is challenging, in particular with respect to \textit{individual} decision-making.
Practitioners frequently struggle to compare training runs due to the inherent stochasticity in algorithms arising from random dithering exploration, environment transition noise, and stochastic gradient updates to name a few.
Traditional analytical approaches, such as replicator dynamics, oft rely on mean-field approximations to remove stochastic effects, but this simplification, whilst able to provide general overall trends, can lead to dissonance between analytical predictions and actual agent realisations.
We propose modelling MARL training as a \textit{coupled stochastic dynamical system}, capturing both agent interactions and environmental characteristics.
Leveraging tools from dynamical systems theory, we pragmatically analyse the stability and sensitivity of agent behaviour, which are key dimensions for their practical deployments, for example, in presence of strict safety requirements, while preserving the stochastic nature of practical MARL, and in doing so explore methods to enact better control.
This framework allows us to rigorously study the inherent stochasticity of MARL, providing a deeper understanding of system behaviour for the design and control of multi-agent learning processes.
}

\begin{document}

\makeCover  
\maketitle  

\begin{abstract}

Analysing learning in Multi-Agent Reinforcement Learning (MARL) environments is challenging, in particular with respect to \textit{individual} decision-making.
Practitioners frequently struggle to compare training runs due to the inherent stochasticity in algorithms arising from random dithering exploration, environment transition noise, and stochastic gradient updates to name a few.
Traditional analytical approaches, such as replicator dynamics, oft rely on mean-field approximations to remove stochastic effects, but this simplification, whilst able to provide general overall trends, can lead to dissonance between analytical predictions and actual agent realisations.
We propose modelling MARL training as a \textit{coupled stochastic dynamical system}, capturing both agent interactions and environmental characteristics.
Leveraging tools from dynamical systems theory, we pragmatically analyse the stability and sensitivity of agent behaviour, which are key dimensions for their practical deployments, for example, in presence of strict safety requirements.
This framework allows us to rigorously study the inherent stochasticity of MARL, providing a deeper understanding of system behaviour.
\end{abstract}


\section{Introduction}
Multi-Agent Reinforcement learning (MARL) routinely exhibits abrupt performance shifts, oscillations, and inconsistent game equilibria during training \citep{mazumdar2020policy, goll2024deterministic}.
RL practitioners observe these phenomena across divergent training runs even under identical architectures and hyperparameters, making stability during training and asymptotic performance both difficult to reason about and costly to validate empirically \citep{henderson2018deep}. 
Unstable training can lead to brittle policies that fail under small perturbations \citep{henderson2018deep, agarwal2021deep}. 
In safety critical applications these effects are particularly problematic \citep{garcia2015comprehensive, amodei2016concrete}. 
Assessing a learning system’s stability and sensitivity to parameters provides crucial insight into robustness, generalisation, and long term adaptability.

There is a long tradition of framing learning as a dynamical system: neuroscience advocated for dynamical systems views of agents and environments \citep{favela2020dynamical}; evolutionary game theory models population adaptation via dynamical systems \citep{maynard1982evolution}; and several RL algorithms have dynamical systems based convergence proofs.
In MARL, evolutionary game theory studies population level adaptation via replicator dynamics \citep{maynard1982evolution}. 
However, these studies either (i) focus on \textit{population level} and \textit{deterministic dynamics}, or (ii) treat \textit{environment as a dynamical system} while representing the learning rule only implicitly.
By focusing on deterministic population averages, the stochastic path dependent dynamics of individual parameter updates are obscured, motivating an individual level perspective to understand stability, convergence, and emergent behaviour.

We study decision making in MARL using dynamical systems theory tools from an \textit{individual} or \textit{agent-centric} perspective, modelling each agent’s learning update as a discrete time dynamical system and treat agents and environment as \emph{coupled dynamical systems}. 
This formulation captures the actual update rules used in MARL, including stochasticity, and allows us to analyse stability and sensitivity directly at the level of individual decision making. 
This approach allows us to leverage the mature toolbox of dynamical systems theory, including methods for analysing: stationary (or invariant) distributions \citep{farmer1982information, scargle1989introduction}, limit cycles, quasi-cycles, random and strange attractors, and Lyapunov stability \citep{lyapunov1992general}. 
We provide a pragmatic use of these tools in the context of MARL\footnote{For a more in-depth description of dynamical systems theory, we refer the reader to the existing excellent resources in this area, e.g., \citep{strogatz2024nonlinear}.} with an analysis focus on agent emergent interaction dynamics.
%
%
Concretely, the contributions of this paper can be summarised as follows:
\begin{itemize}
    \item We formalise individual MARL learning updates as coupled discrete time dynamical systems in parameter space, covering both deterministic and stochastic settings.
    \item We introduce a practical methodology for diagnosing stability and sensitivity in deep MARL using tools from dynamical systems theory, even when closed form analysis is impossible.
    \item We empirically demonstrate the approach across multiple MARL games and learning algorithms, analysing regime transitions and hyperparameter sensitivity.
\end{itemize}

\section{Background}
Multi-agent reinforcement learning (MARL) has achieved strong empirical results across competitive multiplayer games \citep{vinyals2019grandmaster, berner2019dota}, cooperative and competitive social dilemmas \citep{leibo2017multi,AHM20:Partner,AGHM21:Cooperation,DagMus25:Investigating}, and even environmental policy derivation \citep{zhang2022ai, RMP25:multiagent, rudd2025crafting}.
Despite these successes, MARL remains substantially more challenging to train and analyse than its single-agent counterpart. 
Non-stationarity induced by concurrent learners often produces oscillatory behaviours, instability, and brittle dependence on hyperparameters, complicating reproducibility \citep{bloembergen2015evolutionary}.  
MARL guarantees convergence only under restrictive assumptions such as of a zero-sum game \citep{hussain2023beyond} or in the mean field limit \citep{yang2018mean}. 
As a result, a general theoretical understanding of MARL convergence remains elusive.

\textbf{Evolutionary Game Theory \& Replicator Dynamics.} 
Evolutionary Game Theory provides a population-level perspective on adaptation over time via Replicator Dynamics \citep{maynard1982evolution, hofbauer1998evolutionary}, where the agent centric system is reformulated as an aggregate level dynamical system that denotes the evolving community share of agent strategies.
MARL updates and replicator dynamics are strongly linked: the population share of each strategy can be related to the probability of taking a certain action, as if viewing through the lens of a single agent \citep{bloembergen2015evolutionary}.
For example, Cross learning \citep{cross1973stochastic} admits a formal correspondence to replicator dynamics in normal-form games, converging in the continuous time limit \citep{borgers1997learning}, with extensions to certain stochastic and sequential games \citep{hennes2009state, galstyan2013continuous, hennes2020neural}. 
However, there is a disconnect between replicator dynamics theory and practical MARL implementations: classical replicator dynamics assumes infinite populations and omits key algorithmic features such as exploration, bootstrapping, and function approximation that are inherent sources of stochasticity.
Consequently, replicator dynamics provides powerful insights but fails to capture finite agent effects and stochastic learning.

\textbf{Stochastic and Finite-Population Evolutionary Dynamics.}
To bridge idealised infinite-population models and the noisy reality of multi-agent systems, evolutionary game theory increasingly incorporates finite-population effects and stochasticity. 
Finite-population evolutionary dynamics capture demographic noise and drift \citep{nowak2004evolutionary}, since finite-agent constraints can disrupt deterministic attractors causing transitions between equilibria that classical replicator dynamics would classify as completely stable. 
Similarly, evolutionary dynamics with stochastic shocks \citep{fudenberg1992evolutionary, imhof2005long} evaluate how environmental volatility perturbs payoffs and dictates long-term equilibrium selection.
However, a fundamental limitation of both these approaches is their strict reliance on population-level frequency space, which tracks macro-level proportions of strategies across a generalised pool. 
This aggregate abstraction fails to map onto individual, trajectory-dependent parameter updates of learning agents, ignoring critical algorithmic mechanisms like bootstrapping, experience replay, or neural function approximation. 
Furthermore, analytically solving their underlying continuous-time stochastic differential equations becomes mathematically intractable as strategies or states scale, rendering these frameworks unusable for deep MARL.

\textbf{Other Deterministic Methods.} 
Alternatively, \citet{barfuss2019deterministic} look at deriving the deterministic limit of three common RL algorithms - Q-learning, SARSA, and actor critic learning.
By separating the adaption timescale (learning) from the interaction timescale (game dynamics) they map stochastic update rules found in traditional RL algorithms (e.g., $\epsilon$-greedy exploration) to continuous deterministic flows \citep{barfuss2019deterministic}. 
Thus enabling them to leverage deterministic dynamical systems theory to understand the attractors of the systems.
However, adjusting algorithms so that they act in deterministic ways reduces much of their performance as well as the ability to scale to larger state spaces with function approximation.
What is gained in theoretical predictive performance comes at the cost of scalable implementations.

\textbf{Chaos in MARL.}
One of the major difficulties in MARL is the prevalence of chaotic dynamics, which makes proving convergence particularly challenging. 
For example, \citet{sato2002chaos} demonstrate how replicator dynamics can generate complex orbits as well as chaotic attractors and \citet{galla2013complex} and \citet{sanders2018prevalence} show that two player and many player games respectively can exhibit fixed points, limit cycles, or chaotic behaviour.
These findings raise the fundamental question of whether fixed points are consistently attainable in such systems at all, or whether approximations and assumptions used to enforce convergence inadvertently strip away the very dynamics that could be essential for capturing the richness of agent interactions. 
In this sense, the ``chaos'' observed in MARL is a fundamental characteristic, posing significant challenges for both theory and practice.

\section{Our Approach}
In this work we focus on understanding individual decision making of MARL agents, an agent-centric rather than population level viewpoint.   
Our perspective is therefore complementary to population based approaches: rather than modelling population averages, we directly analyse the learning dynamics of individual agents.

\subsection{Definitions}
\textbf{Dynamical Systems.} 
A dynamical system is a mathematical framework for describing the evolution of a system over time, formally consisting of a state space $X$ together with an evolution rule $\phi_t$ that describes how the state changes. 
The state space is typically in $\mathbb R^n$ or a manifold, and the evolution rule can be either continuous or discrete in time as well as either deterministic or stochastic.
Continuous time deterministic dynamics are expressed by an Ordinary Differential Equation (ODE) $\dot x = f(x(t)), \; x(0) = x_0$, where $f : X \rightarrow \mathbb R^n$ is a vector field, and the solution is the flow $\phi_t(x_0)$ that represents a trajectory starting from the initial state.
Discrete time deterministic dynamics are described by an iterated map $x_{t+1} = F(x_t), \; x_0 \in X$, where $F:X \rightarrow X$ is a transformation and the system evolves by iteration of $F$.

One can visualise the state space (aka phase space) as a landscape of possible states that the dynamical system evolves through. 
Trajectories (realisations) of the system emit structure in the phase space, known as a phase plot/portrait.
Attractors are locations in phase space the system moves towards regardless of initial conditions, indicating asymptotic behaviour.
There are three main types of attractors: fixed points - point locations in which the system converges, limit cycles - attractors the system periodically loops around, strange attractors - bounded region of phase space where irregular system behaviour leads to a fractal structure.

Coupled dynamical systems are comprised of multiple dynamical systems that have cross/coupling terms within $f$ or $F$, such as the bidirectionally coupled Logistic Map \citep{sugihara2012detecting}:
\begin{align}
    x_{t+1} &= x_t(r_x - r_x x_t - \beta_{xy} y_t) \nonumber \\
    y_{t+1} &= y_t(r_y - r_y y_t - \beta_{yx}x_t),
\end{align}
where $r_x, r_y, \beta_{xy}, \beta_{yx}$ are parameters that adjust the chaotic behaviour and coupling strength.

\noindent \textbf{Markov Decision Process/Markov Game.} 
RL utilises Markov Decision Processes (MDP), defined by the tuple $\left< \mathcal{S}, \mathcal{A}, P, R, \gamma \right>$. 
Where $\mathcal{S}$ is the set of states, $\mathcal{A}$ the set of actions, $P(s' \mid s, a)$ the transition probability of moving to state $s'$ after taking action $a$ in state $s$, $R(s,a)$ the reward function, and $\gamma \in [0,1)$ a discount factor.
At each timestep $t$, the agent observes $s_t \in \mathcal{S}$, chooses $a_t \in \mathcal{A}$, receives reward $r_t = R(s_t,a_t)$, and transitions to $s_{t+1} \sim P(\cdot \mid s_t, a_t)$.
Generalising for multiple agents, the MDP becomes a stochastic game or Markov Game (MG), a multi-agent MDP with $N$ agents defined by the tuple $ \left< \mathcal{S}, \{\mathcal{A}^i\}_{i=1}^N, P, \{R^i\}_{i=1}^N, \gamma) \right>$.
Where $\mathcal{S}$ is the shared state space, $\mathcal{A}^i$ is the action space of agent $i$, $P(s' \mid s, a^1, \dots, a^N)$ is the transition probability dependent on the joint action $\mathbf{a} = (a^1, \dots, a^N)$, and $R^i(s, \mathbf{a})$ is the reward function for agent $i$. 
Unlike the single-agent case, where the environment is stationary under a fixed policy, multi-agent systems are inherently non-stationary since the dynamics depend on the evolving policies of all agents. 

\textbf{Single-Agent RL as a Dynamical System.} 
Consider a discrete time dynamical system for a singular agent. 
We initially assume a fully deterministic framework, and subsequently identify and discuss the sources of stochasticity:
\begin{equation}
    s_{t+1} = f(s_t, a_t),    
\end{equation}
where $a_t$ is the agent action, $s_t$ the state at time step $t$, and $f(\cdot, \cdot)$ our deterministic transition function that is affected by agent actions.
Similarly to \citet{beer1995dynamical} we model a coupled dynamical system of the learning RL based agent and the environmental dynamical system:
\begin{equation}
    \theta_{h+1} = g(\theta_h, \mathbf s_h, \mathbf a_h), 
\end{equation}
where $h$ is a learning update step, relating to a certain number of time steps $t$. 
$\bm \theta_h$ (a scalar or vector) is the agent state at update $h$, perhaps the model parameters, and an update depends on the trajectory of states and actions gained in that update window. 
Actions are generated from a policy $\pi$ using the environment state $s_t$ and the agent state $a_t=\pi(x_t; \theta_h)$
\textit{We can understand the stability of the function $g$ by analysing what kind of attractor it settles upon.}
This is the simplest definition form, but can be extended to incorporate practical implementations such as replay buffers, target networks, or off policy learning, where the samples for updates may originate from different behavioural policies.

\textbf{MARL as a Dynamical System.} 
For clarity we assume just two agents in this system but the definition can be expanded to general agents.
Upon introducing multiple agents our discretised transition dynamics become:
\begin{equation}
    s_{t+1} = f(s_t, a_t^1, a_t^2),    
\end{equation}
where the superscript identifies the agent. 
Creating a coupled dynamical system between the updates of individual agents as their trajectories are dependent on the updates of other agents, defined as:
\begin{align}
    \theta_{h+1}^1 &= g^1(\theta_h^1, \mathbf s_h, \mathbf a_h^1) \nonumber \\
    \theta_{h+1}^2 &= g^2(\theta_h^2, \mathbf s_h, \mathbf a_h^2) ,
\end{align}
where an agent updates its internal representation given a set of historical states and its own actions. 
Different MARL algorithms change $g^i$ (e.g., independent learners or opponent modelling), altering the coupling strength/structure. 
Additional variables can be added creating further coupling, such as opponent actions or other feature representations used for agent updates.
In the following, we focus on the simplest case where an agent makes decisions about the system only from global state information and its own action, defined in agent parameter space:
\begin{align}
    \theta_{h+1}^1 &= g^1 \left(\theta_h^1, f \left(s_t, \pi_{\theta^1}(s_t), \pi_{\theta^2}(s_t) \right), \pi_{\theta^1} (s_t) \right) \nonumber \\
    \theta_{h+1}^2 &= g^2 \left(\theta_h^2, f \left(s_t, \pi_{\theta^1}(s_t), \pi_{\theta^2}(s_t) \right), \pi_{\theta^2}(s_t) \right).
\label{eqn:main_dynamical_system_two_agent}
\end{align}
Our goal is to \textit{measure and compare} stability and sensitivity of $g$ across MARL algorithms, games, and hyperparameters, beyond only fixed point existence.

\textbf{The Impacts of Stochasticity.} 
Equation~\ref{eqn:main_dynamical_system_two_agent} presents the simplest coupled dynamical system for a fully deterministic setting.
In practical settings, multiple sources of stochasticity arise within the system. 
Traditionally, these stochastic elements have been approximated deterministically to enable analytical tractability. 
In contrast, we explicitly model these sources of randomness to more faithfully capture the behaviour of MARL agents as a coupled stochastic dynamical system:
\begin{align}
    \theta_{h+1}^1 &= g^1 \left(\theta_h^1, f \left(s_t, \pi_{\theta^1}(s_t) + \xi_t, \pi_{\theta^2}(s_t) + \xi_t \right) + \eta_t, \pi_{\theta^1} (s_t) + \xi_t \right) + \zeta_h \nonumber \\
    \theta_{h+1}^2 &= g^2 \left(\theta_h^2, f \left(s_t, \pi_{\theta^1}(s_t) + \xi_t, \pi_{\theta^2}(s_t) + \xi_t \right) + \eta_t, \pi_{\theta^2}(s_t) + \xi_t \right) + \zeta_h,
\end{align}
where $\xi_t$ represents stochasticity in exploration strategies or policy sampling, $\eta_t$ the environment transition noise, and $\zeta_h$ the stochastic gradient updates.
Below, we present a version that subsumes all sources of stochasticity into a single combined stochastic effect term $\nu_h$:
\begin{align}
    \theta_{h+1}^1 &= g^1 \left(\theta_h^1, f \left(s_t, \pi_{\theta^1}(s_t), \pi_{\theta^2}(s_t) \right), \pi_{\theta^1} (s_t) \right) + \nu_h \nonumber \\
    \theta_{h+1}^2 &= g^2 \left(\theta_h^2, f \left(s_t, \pi_{\theta^1}(s_t), \pi_{\theta^2}(s_t) \right), \pi_{\theta^2}(s_t) \right) + \nu_h.
    \label{eqn:accumulated_noise_two_agent}    
\end{align}

\textbf{Comparison to Replicator Dynamics.} 
We compare our agent centric perspective to the population level replicator dynamics formulated as a continuous time dynamical system:
\begin{equation}
    \dot u_i = u_i \left[w_i(\mathbf u) - \bar w(\mathbf u) \right],
\end{equation}
where $\mathbf u = (u_1,u_2,...,u_n)$ represents the population state vector quantifying the percentage of the population that belong to each of the $n$ strategies, $w_i(\cdot)$ is the fitness function of a specific strategy, and $\bar w(\cdot)$ is the average fitness of the total population.
As the $\bar w(\cdot)$ term is an expectation over the whole population, this averages out many of the stochastic effects.

\subsection{Dimensions of our Analysis}
Equation \ref{eqn:accumulated_noise_two_agent} provides a general definition of a coupled dynamical system for two agents, which during learning generates a phase portrait in $(\theta^1,\theta^2)$ for any combination of environment and MARL agents.
Leveraging stochastic dynamical systems theory, we analyse system behaviour along three key dimensions:

\begin{itemize}
    \item \textit{Stability analysis:} Understanding the asymptotic performance of a MARL system allows us to rigorously validate the resulting game equilibria.
    \item \textit{Sensitivity analysis:} Once the phase-space structure can be analysed, we can assess how parameter changes affect stability, providing insight into the system's sensitivity.
    \item \textit{Control:} By quantifying stability and sensitivity in the coupled MARL dynamical system, we can inform strategies for improved control.
\end{itemize}

Our framework is agent-centric and data driven: rather than assuming deterministic structure, we analyse realised stochastic trajectories under arbitrary algorithm environment pairings. 
In this setting, classical notions (e.g. fixed points, attractors, stability) must be interpreted in a stochastic sense, as process noise perturbs trajectories and precludes strict invariance.
Formally, each agent's update follows a Markov process $\theta_{h+1} = g(\theta_h, s_t, \bm \pi) + \nu_h$, and behaviour is characterised through a stationary distribution $\rho^\theta$ that satisfies $\theta_h \sim \rho^\theta \Rightarrow \theta_{h+1} \sim \rho^\theta$.

Although closed-form solutions for $\rho^\theta$ are generally intractable, ergodic approximations can be obtained via long-run simulations or multiple initialisations. 
The geometry of $\rho^\theta$ reveals qualitative system behaviour.
Concentrated mass indicates a fixed point, while sustained oscillations appear as a ``smeared" ring, characteristic of quasi-cycles induced by stochasticity.
Quasi-cycles arise when deterministic dynamics would converge to a fixed point or limit cycle, but stochastic perturbations sustain oscillatory behaviour.
To make this concrete, we introduce four diagnostics:

\begin{itemize}
    \item \textit{Stationary (invariant) distributions \citep{farmer1982information, scargle1989introduction}:} Probability distribution over $\theta$. If the Frobenius norm $\|\Sigma\|_F$ of the covariance of $\rho^\theta$ is low it indicates fixed point convergence.

    \item \textit{Lyapunov exponents \citep{lyapunov1992general}:} They quantify how quickly two trajectories diverge or converge, a hallmark of stability or chaos. For trajectories $\theta_h$ and $\theta_h'$ starting $\epsilon$ apart, $\lambda = \lim_{h \rightarrow \infty} \frac{1}{h} \log \frac{||\theta_h - \theta_h'||}{||\theta_0 - \theta_0' ||}$. Negative exponents suggest convergence to a noisy fixed point, near-zero values indicate cycles or neutral stability, and positive exponents are evidence of chaos. 
    
    \item \textit{Recurrence plots \citep{eckmann1995recurrence}:} Dynamical systems over time can visit states recurrently, visualising the pattern of these revisits indicates ordered, periodic, or chaotic behaviour. Defined as a binary matrix $R_{ij}$ where $R_{ij} = 1 \quad \text{if} \; \; ||\theta_i - \theta_j|| < \epsilon, \quad \text{else} \; \; 0$.  
    
    \item \textit{Fractal dimensions of attractors \citep{theiler1990estimating}:} Chaotic attractors often have fractal geometry occupying a fractional dimension between integer values. Using states $\theta_i$, calculating: $C(r) = \frac{1}{N^2} \sum_{i,j} \bm 1\{ || \theta_i - \theta_j || < r \}$ denotes the probability two points are within distance $r$. For small $r$, $C(r) \sim r^{D_2}$ where $D_2$ is the correlation dimension. $D_2 \approx 0$ is a fixed point, $D_2 \approx 1$ a limit cycle, and $D_2$ non-integer $>1$ a fractal attractor (signal of chaos).
\end{itemize}

Please refer to Appendix \ref{appendix:stability} for details on how these diagnostics are computed.

\section{Experimental Settings}
We evaluate all methods under a shared experimental protocol across four canonical stateless repeated games: Prisoner’s Dilemma, Matching Pennies, Stag Hunt, and Chicken, and a larger multi-state cooperative environment, Simple Spread from Multi Particle Environments (MPE) \citep{mordatch2017emergence}. 
The matrix games capture standard social dilemma, zero-sum, coordination, and competitive structures, respectively, while Simple Spread provides a high-dimensional setting requiring sustained coordination between three agents.
We use two or three independent learner algorithms that closely align with replicator dynamics.
Tabular Q-learning approximates replicator dynamics in stateless games in the continuous time, infinitesimal step-size limit if it uses Boltzmann (Softmax) exploration \cite{bloembergen2015evolutionary}.
Policy gradient (REINFORCE with baseline) reduces to replicator dynamics in the mean-field limit, since subtracting the baseline corresponds to subtracting the average payoff \citep{bernasconi2025evolutionary}.
To scale to Simple Spread, we use Independent Deep Q-Networks (IDQN) \citep{tampuu2017multiagent}, introducing function approximation and additional stochasticity.
Appendix \ref{appendix:environments} presents detailed descriptions of these environments.

\begin{figure}[t]
\captionsetup{skip=0pt}
    \centering
    \includegraphics[width=1\textwidth]{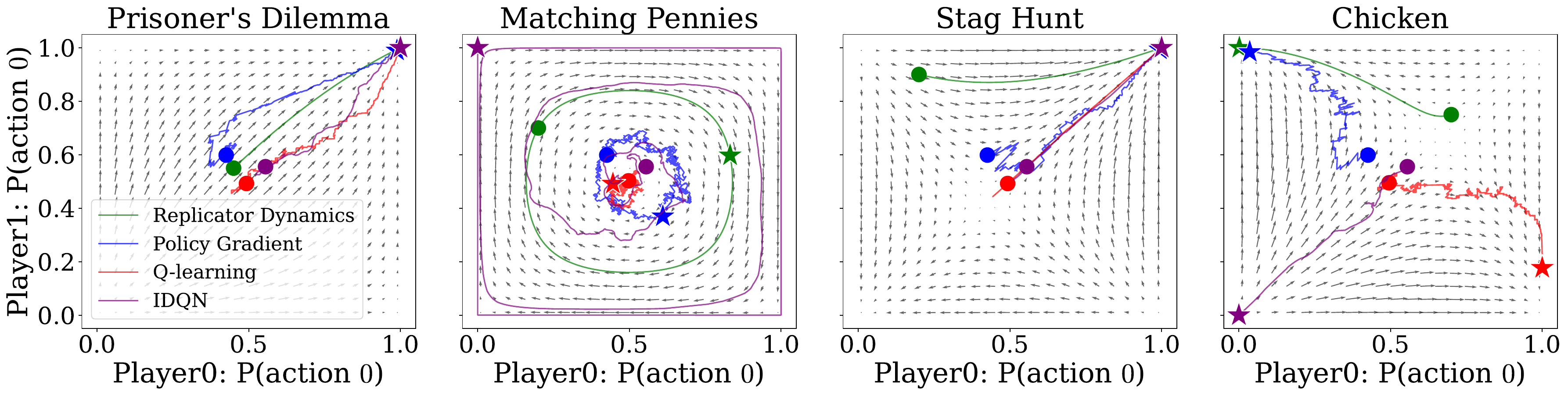}
        \caption{
            Comparison between replicator dynamics and Policy Gradient, Tabular Q-learning, and IDQN with Boltzmann exploration.
            Replicator dynamics are represented by the analytical vector field and an arbitrary initial condition realisation.
            Realisations start from circles and end at stars.
        }
    \label{fig:exp_1}
    \vspace{-10pt}
\end{figure}

\section{Understanding Individual Decision-Making in MARL}
\subsection{Stability Analysis}
Stability analysis in MARL concerns asymptotic behaviour and its relation to game equilibria. 
Evolutionary game theory characterises equilibria as fixed points of deterministic dynamics \citep{hofbauer1998evolutionary}. 
When replicator dynamics apply, the induced flow admits a closed-form vector field.
As our approach is data-driven, we visualise empirical trajectories in parameter space (policy traces). 
Figure \ref{fig:exp_1} compares phase portraits with policy traces from Q-learning, Policy Gradient, and IDQN with Boltzmann exploration, where parameters correspond to action probabilities. 
With sufficiently small learning rates, Q-learning and Policy Gradient approximate replicator flow. 
In contrast, IDQN with Boltzmann exploration introduces additional stochasticity from function approximation, gradient updates, and target networks, yielding behaviours such as expanding cycles in Matching Pennies and drift toward defection in Chicken.

\begin{figure}[t]
    \captionsetup{skip=0pt}
    \centering
    \includegraphics[width=1\textwidth]{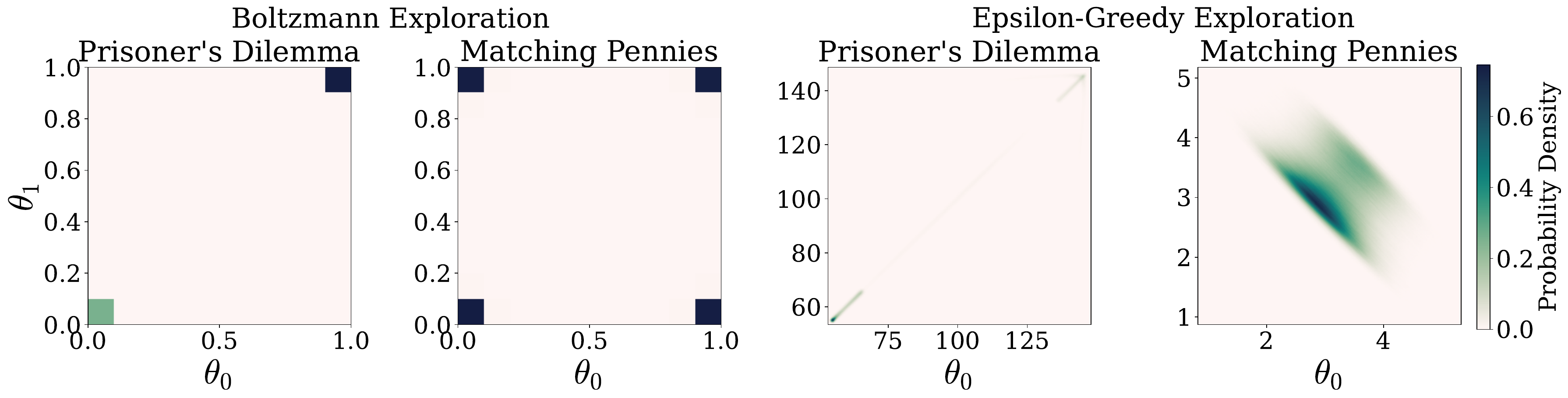}
    \caption{Stationary distributions calculated from realisations of training two IDQN agents.
    The two left figures are agents using Boltzmann exploration; the parameters can be interpreted as probabilities of taking action $0$.
    On the Boltzmann plots bin counts are intentionally very low so it is clear where the stationary distribution has density.
    The two right figures are agents using $\epsilon$-greedy exploration.
    Parameters cannot be interpreted as action probabilities which is why their scale is larger.
    }
    \label{fig:mp_pd_stationary_dist}
    \vspace{-10pt}
\end{figure}

Figure \ref{fig:mp_pd_stationary_dist} shows stationary distributions for Prisoner’s Dilemma and Matching Pennies under IDQN with Boltzmann and $\epsilon$-greedy exploration.
Under Boltzmann exploration, the stationary distribution largely concentrates near the sink points predicted by replicator dynamics in Figure \ref{fig:exp_1}.
In Prisoner’s Dilemma this appears as strong concentration near mutual defection, while Matching Pennies exhibits spread across parameter space consistent with cycling behaviour, seen in Figure \ref{fig:exp_1}.
Some mass in the bottom left indicates that for some initial conditions convergence does not match replicator dynamics.
Under $\epsilon$-greedy exploration, dynamics become more varied: Prisoner’s Dilemma converges to one of two fixed points, whereas Matching Pennies displays an approximate limit cycle, consistent with quasi-cyclic dynamics.

\begin{figure}[htbp]
    \captionsetup{skip=0pt}
    \centering
    \includegraphics[width=0.55\textwidth]{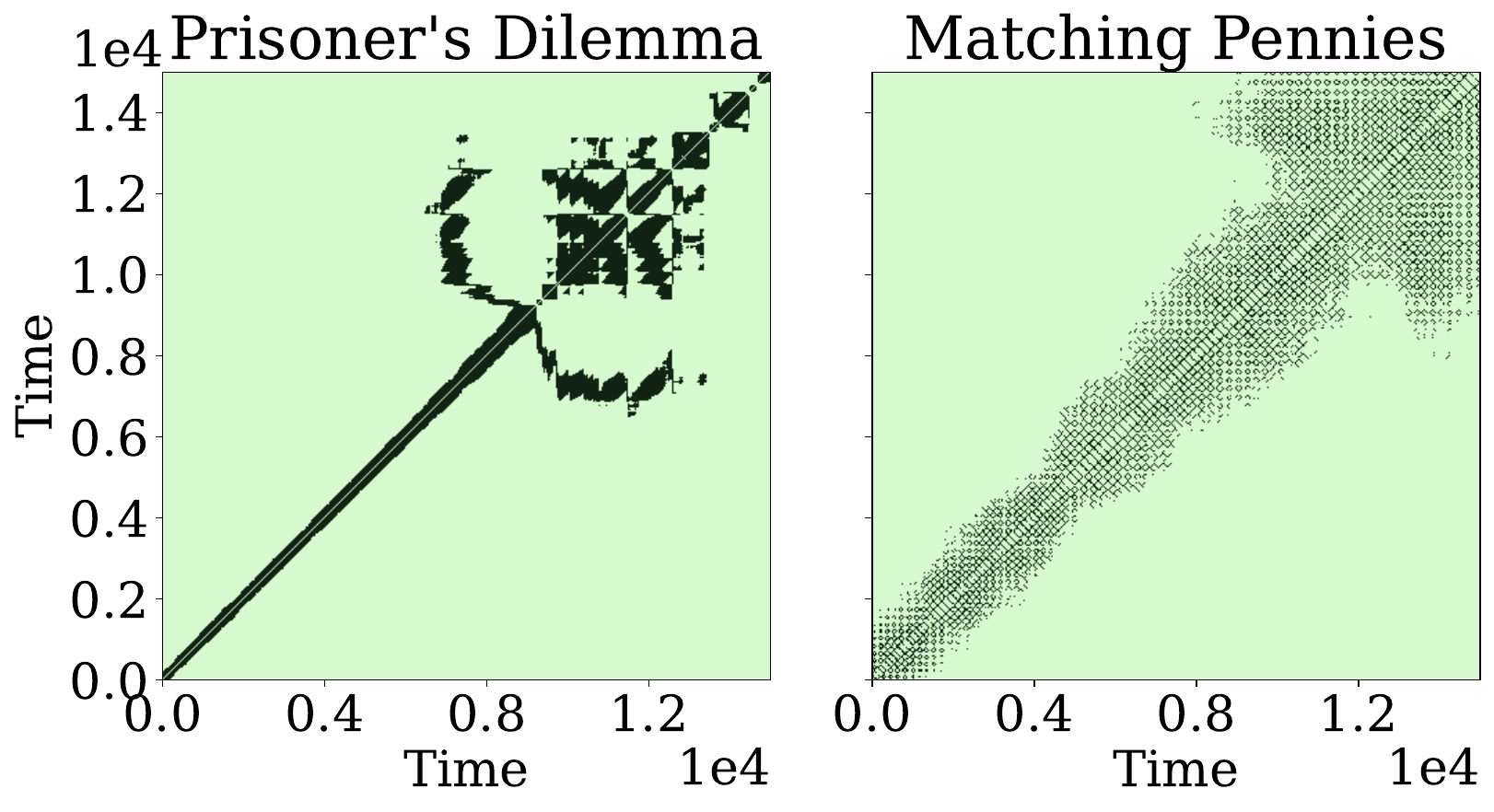}
    \caption{
    Recurrence plots from a realisation of training two IDQN agents, indicating when the coupled dynamical system of all agents visits the same area in phase space at the time on the $x$-axis and $y$-axis.
    Intuitively, this means locations are marked when $\theta_i \approx \theta_j$ when $i=x, j=y$, with identity band masked out as it is always recurrent when $i=j$ so adds no extra information.
    }
    \label{fig:mp_pd_recurrence_plot}
    \vspace{-10pt}
\end{figure}

Further quantitative evidence (Table \ref{tab:all_envs_diagnostics}) supports these observations. 
In Prisoner’s Dilemma, $\lambda_{\max} \approx 0$ and fractal dimension $D_2 \approx 0$ indicate convergence to a stable fixed point. 
In contrast, Matching Pennies exhibits positive $\lambda_{\max}$ and larger $D_2$, consistent with cyclical behaviour.
Figure \ref{fig:mp_pd_recurrence_plot} presents recurrence plots for both games (with the identity line masked). 
In Prisoner’s Dilemma, long diagonal bands parallel to the identity line dominate, indicating deterministic and predictable dynamics (a stochastic process exhibits almost none) \citep{marwan2007recurrence}. 
The pattern here suggests the system follows a regular, repeating path through its state space.
Matching Pennies instead shows regularly spaced diagonals, characteristic of periodic motion, along with arc like structures associated with quasi-periodicity \citep{facchini2007curved}.

\begin{table}[t]
    \captionsetup{skip=0pt}
    \caption{Diagnostics for stochastic fixed points, limit/quasi-cycles, and chaotic attractors with $95\%$ confidence intervals over 8 random seeds.}
    \centering
    \begin{tabular}{lccccc}\toprule
        \textbf{Environment} & $\|\Sigma\|_F$ & $\lambda_{\max}$ & $D_2$ \\
        \midrule
        Prisoners' Dilemma & $0.102 \pm 0.041$ & $\approx 0$ & $0.438 \pm 0.042$ \\
        Matching Pennies & $2.351 \pm 0.103$ & $0.039 \pm 0.002$ & $1.154 \pm 0.075$ \\
        Stag Hunt & $0.553 \pm 0.068$ & $\approx 0$ & $0.628 \pm 0.050$ \\
        Chicken & $0.118 \pm 0.022$ & $\approx 0$ & $0.760 \pm 0.034$ \\
        Simple Spread & $0.956 \pm 0.015$ & $\approx 0$ & $0.441 \pm 0.064$ \\
    \bottomrule
    \end{tabular}
    \label{tab:all_envs_diagnostics}
    \vspace{-10pt}
\end{table}

\begin{figure}[htbp]
    \captionsetup{skip=0pt}
    \centering
    \includegraphics[width=0.3\textwidth]{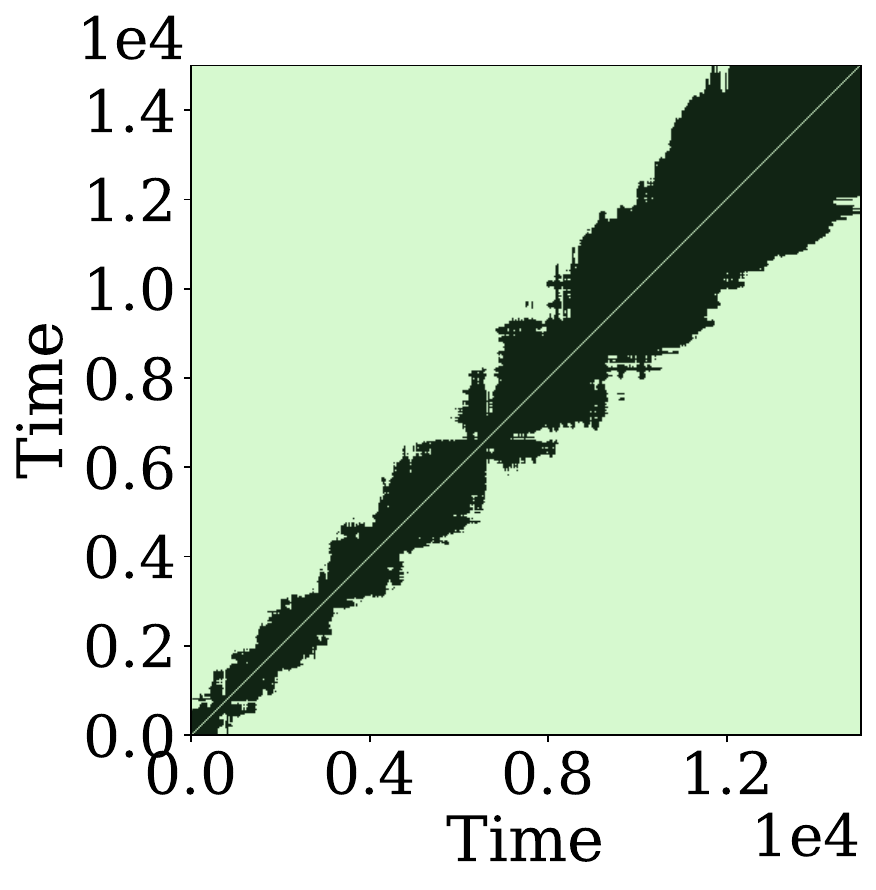}
    \caption{
    Recurrence plot from a realisation of training three IDQN agents in Simple Spread.
    }
    \label{fig:simple_spread_recurrence}
    \vspace{-10pt}
\end{figure}

Thus far, we have considered settings amenable to direct visualisation. 
As the number of agents or parameters grows, phase portraits become impossible. 
Although empirical diagnostics remain well-defined, intuitive inspection is hindered. 
High-dimensional systems typically require projection (e.g., via Principal Component Analysis \citep{hotelling1933analysis}), incurring information loss.
Recurrence plots offer an alternative, providing a two-dimensional representation of the dynamics \citep{eckmann1995recurrence}, capturing structural properties such as determinism and convergence (Figure \ref{fig:simple_spread_recurrence}).
Pairing the recurrence plots with Table \ref{tab:all_envs_diagnostics} for the higher-dimensional cooperative setting of Simple Spread, the diagnostics reveal dynamics between deterministic convergence and cyclical instability. 
The relatively large Frobenius norm ($||\Sigma||_F = 0.956$) indicates moderate parameter variance among the three agents. 
Furthermore, the near-zero maximum Lyapunov exponent and the correlation dimension ($D_2 = 0.441$) suggest that the system settles into a state of stability. 
Rather than collapsing to a strict fixed point ($D_2 \approx 0$), the agents' joint policy forms a quasi-cycle driven by the inherent stochasticity of IDQN's function approximation. 
Figure \ref{fig:simple_spread_recurrence} corroborates these quantitative findings: exhibiting a broad, dense diagonal band with diffuse edges stemming from exploration noise and multi-agent non-stationarity in contrast to sharp diagonal lines seen in deterministic settings.

To summarise, analysing the MARL system as an agent-centric coupled dynamical system enables a practitioner to clearly understand the stability and asymptotic performance of an agent's \textit{true} behaviour.
Further, our quantitative and qualitative insights are able to scale to environments and agent numbers that can't be easily visualised, opening the door for analysis in any environment.
When working with classical state based games, the insights from replicator dynamics generally align with our findings, indicating the usefulness of a paired approach for understanding MARL stability.

\begin{figure}[t]
    \captionsetup{skip=0pt}
    \centering
    \includegraphics[width=1\textwidth]{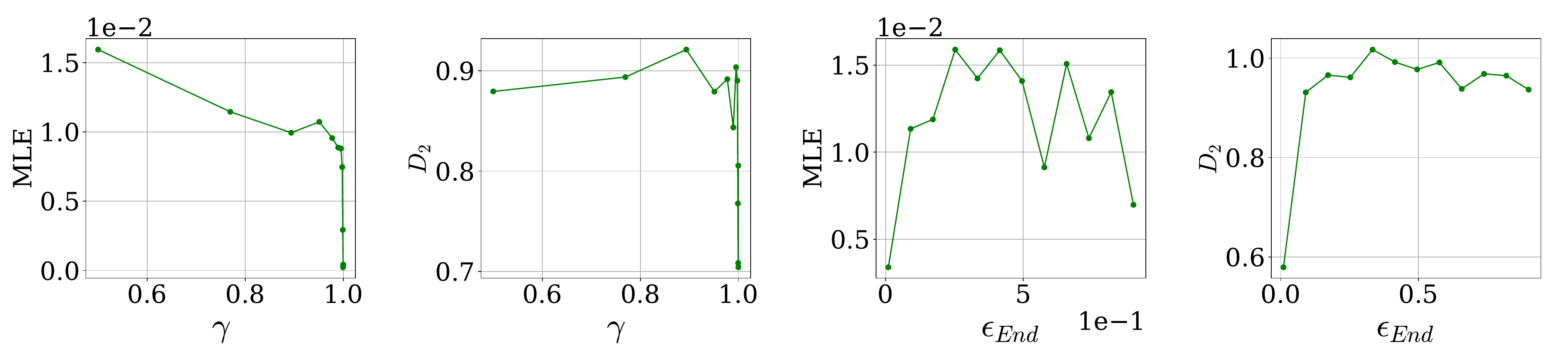}
    \caption{
    Varying $\gamma$, the discounting parameter in IDQN, and $\epsilon_\text{End}$ the end value for $\epsilon$-greedy exploration in IDQN, to understand their impact on the coupled dynamical system attractor via the Maximum Lyapunov Exponent (MLE) and fractal dimension $D_2$ over 8 random seeds.
    }
    \label{fig:mp_sensitivity}
    \vspace{-10pt}
\end{figure}

\subsection{Sensitivity Analysis}
\begin{figure}[t]
    \captionsetup{skip=0pt}
    \centering
    \includegraphics[width=1\textwidth]{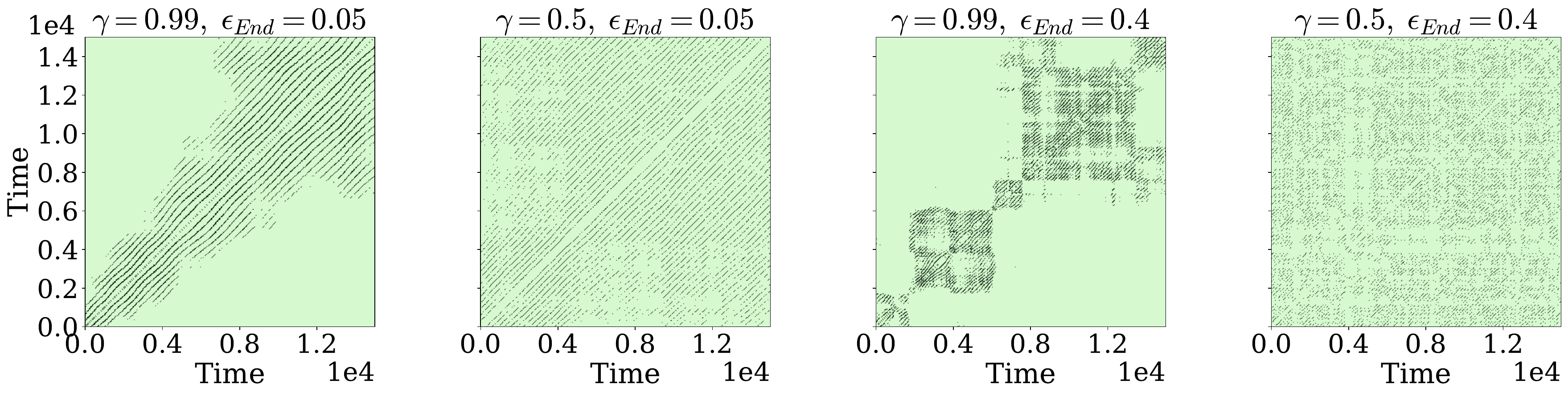}
    \caption{
    Recurrence plot calculated from a realisation of training two IDQN agents in Matching Pennies.
    All four combinations of $\gamma = \{0.5, 0.99\}$ and $\epsilon_\text{End}=\{0.4,0.0\}$ are presented.
    }
    \label{fig:sensitivity_recurrences}
    \vspace{-10pt}
\end{figure}

We analyse how MARL dynamics change under variations in hyperparameters and learning rules, focusing on topological shifts in phase space induced by different choices of $g$. 
We sweep the discount factor $\gamma$ and terminal exploration rate $\epsilon_{\text{End}}$ in IDQN (with $\epsilon=0.9$ and exponential decay), estimating the maximal Lyapunov exponent and correlation dimension $D_2$ from post-burn-in policy traces across seeds (Figure \ref{fig:mp_sensitivity}).
Increasing $\gamma$ suppresses cycling in Matching Pennies as $\gamma \to 1$. 
Setting $\epsilon_{\text{End}}=0$ removes exploration noise, yielding convergence to a fixed point ($\lambda_{\max}, D_2 \to 0$). 
Interestingly, small positive $\epsilon_{\text{End}}$ induces cyclical behaviour, while larger values reduce structure, suggesting convergence toward smaller radius limit cycles or quasi-cycles compared to the case with $\epsilon_{\text{End}} \approx 0.4$.
Recurrence plots (Figure \ref{fig:sensitivity_recurrences}) corroborate this transition: the far left figure shows a clear limit cycle with determinism, the centre left a quasi-cycle as the spread of diagonal lines indicates higher stochasticity, the centre right has hallmarks of the beginnings of chaos, and finally the far right is almost purely stochastic with very faint structure.
These sensitivity experiments reveal not only the phase space \textit{shape}, but also \textit{why} it assumes that shape under different hyperparameter settings, providing a principled connection between tuning choices and long-term dynamical behaviour.

\subsection{Control}
In the previous sections, having characterised stability and sensitivity through the lens of dynamical systems theory, we propose closing the loop by using these metrics to actively shape MARL behaviour. 
Rather than serving purely as post-hoc diagnostics, quantities such as the maximal Lyapunov exponent and correlation dimension can define control objectives (e.g., suppressing cycling by driving $\lambda_{\max}<0$ and $D_2 \to 0$). 
One possible approach is stability aware MARL, where dynamical metrics act as pseudo-rewards or guide meta-learning of hyperparameters toward desired phase space properties. 
Such schemes complement reward maximisation by steering learning toward regimes that are not only high performing but also stable and predictable.

\section{Implications}
Our central aim is to analyse MARL at the level where decisions occur: individual learning updates. 
Population based approaches (e.g., replicator dynamics) offer aggregate descriptions but smooth out heterogeneity, stochasticity, and algorithmic structure that are central to practical MARL. 
An agent-centric dynamical systems perspective instead models the coupled update dynamics that drive learning and interaction.
This perspective enables principled analysis of stability and sensitivity under stochasticity and function approximation. 
Tools such as invariant distributions, Lyapunov exponents, and recurrence analysis characterise whether learning converges to fixed points, cycles, or chaotic regimes. 
Practically, stable equilibria correspond to consistent joint policies, while oscillatory or chaotic behaviour reflects persistent non-stationarity.
By examining how hyperparameters or environment perturbations shift these regimes, we obtain measures of robustness and reproducibility. 
Although conducted in parameter space, results map to reward and behavioural outcomes, enabling assessment of whether convergence is desirable. 
Applicable to both tabular and deep MARL, this framework provides diagnostics for instability in high-dimensional settings where mean-field abstractions are intractable, clarifying how design choices shape long-term dynamics.

\section{Conclusion}
In this work, we proposed an agent-centric dynamical systems framework for analysing MARL, modelling individual learning updates as coupled stochastic dynamical systems, enabling stability and sensitivity analysis beyond deterministic population level abstractions. 
Using invariant distributions, Lyapunov exponents, recurrence structure, and fractal dimensions, we demonstrated how regime transitions, oscillations, and instability arise in both tabular and deep MARL, and how hyperparameters shape long-run behaviour. 
Our approach is primarily empirical, relying on finite sample trajectory estimates that may be sensitive to burn-in, dimensionality, and projection effects; scalability to very high dimensional multi-agent systems poses computational challenges; and we do not yet provide formal predictive guarantees linking algorithm design to specific attractor structures. 

Future work proceeds in three directions. 
First, additional tools from dynamical systems theory could broaden stability analysis. 
Second, stochastic invariant distributions could be characterised analytically via the Fokker-Planck equation \citep{risken1989fokker} or estimated using partial differential equation based methods \citep{li2021fourier}, enabling direct calculations of stability and sensitivity, as demonstrated by \citet{leung2023stochastic}.
Third, diminishing sources of stochasticity (e.g., decaying exploration) raise open questions about their impact on attractor structure. 
Viewing learning as a coupled dynamical process provides a principled foundation for analysing and controlling emergent behaviour in multi-agent systems.

\subsubsection*{Broader Impact Statement}
\label{sec:broaderImpact}
The methods and findings presented in this work have the potential to support a wide range of socially beneficial applications. 
For example, improving the stability and robustness of multi-agent learning systems could enhance coordination in domains such as distributed robotics, traffic and resource management, and environmental monitoring, especially in areas where safety is paramount.
Such applications can advance sustainability efforts, climate research, biodiversity protection, and responsible resource management.
We advocate for responsible innovation and encourage the use of this work in ways that prioritise environmental sustainability, social equity, and the public good.

\appendix


\section{Environments}
\label{appendix:environments}

We first consider a set of four stateless matrix games, which serve as canonical benchmarks for analysing multi-agent interaction dynamics.
These environments consist of single-step interactions in which agents select actions simultaneously and receive rewards according to fixed payoff matrices. 
As there is no temporal component or state transition, they provide a controlled setting for examining coordination, competition, and equilibrium behaviour under simplified yet illustrative conditions.
The environments are:
\begin{itemize}
\item \textit{Prisoner’s Dilemma}: a standard social dilemma with a unique Nash equilibrium (defection), though cooperation can arise under repeated interaction \cite{axelrod1981evolution,nowak2006five}.
\item \textit{Matching Pennies}: a zero-sum game with a mixed-strategy equilibrium. 
\item \textit{Stag Hunt}: a coordination game with multiple equilibria (cooperative ``stag'' or risk-dominant ``hare''). 
\item \textit{Chicken}: a cooperative/competitive game where mutual aggression is costly, with two pure-strategy equilibria (one player swerves while the other does not) and a mixed-strategy equilibrium. 
\end{itemize}
The pay-off matrices of these games are reported in the tables below.

\begin{table}[H]
\centering

\begin{minipage}{0.24\textwidth}
\centering
\caption*{Prisoner's Dilemma}
\begin{tabular}{c|cc}
      & C & D \\ \hline
    C & 1,1 & 5,0 \\
    D & 0,5 & 3,3 \\
\end{tabular}
\end{minipage}
\hfill
\begin{minipage}{0.24\textwidth}
\centering
\caption*{Matching Pennies}
\begin{tabular}{c|cc}
      & H & T \\ \hline
    H & 1,-1 & -1,1 \\
    T & -1,1 & 1,-1 \\
\end{tabular}
\end{minipage}
\hfill
\begin{minipage}{0.24\textwidth}
\centering
\caption*{Stag Hunt}
\begin{tabular}{c|cc}
      & S & H \\ \hline
    S & 4,4 & 0,0 \\
    H & 0,0 & 3,3 \\
\end{tabular}
\end{minipage}
\hfill
\begin{minipage}{0.24\textwidth}
\centering
\caption*{Chicken}
\begin{tabular}{c|cc}
      & A & B \\ \hline
    A & -1,-1 & 4,0 \\
    B & 0,4 & 2,2 \\
\end{tabular}
\end{minipage}
\end{table}
\vspace{-10pt}

We also use the \texttt{Simple Spread MPE} environment from the \texttt{JaxMARL} benchmark suite \citep{rutherford2024jaxmarl}. 
In this cooperative multi-agent task, multiple agents must coordinate to cover a set of target landmarks in a shared continuous space. 
Each agent aims to occupy a different landmark while avoiding collisions with other agents, requiring effective spatial coordination and decentralised decision making.
Agents must distribute themselves across landmarks such that each landmark is covered by a different agent while minimising collisions. 
The environment’s partially observable state and the need for coordinated positioning make it a useful testbed for studying cooperation and learning dynamics in MARL.

\section{Dynamical Systems Analysis Techniques: Additional Details}
\label{appendix:stability}
This appendix details the numerical methods used to analyse emergent agent dynamics, implemented via vectorised operations in \texttt{JAX} \citep{jax2023} and \texttt{NumPy} \citep{harris2020array}.

\subsection{Stationary Distribution Estimation}
We estimate the invariant measure of the coupled stochastic learning process empirically from long-run policy trajectories. 
Simulating the system for $n_\text{steps}$ iterations across $n_\text{runs}$ random seeds and discarding a burn-in phase of $n_\text{burn}$ steps, the stationary samples for agent parameters $\bm \theta_h = (\theta_h^1, \theta_h^2)$ at update step $h$ are collected as:
\begin{equation}
\mathcal{S} = \bigcup_{i=1}^{n_\text{runs}} \{\bm \theta_h^{(i)} : h > n_\text{burn}\}.
\end{equation}
For scalar two-agent parameters, these form a normalised 2D empirical density $\hat{p}(\theta^1,\theta^2)$; for higher-dimensional spaces or more agents, dimensionality reduction is applied prior to visualisation.

\subsection{Covariance Analysis and Frobenius Norm}
For a multivariate trajectory $\bm{\theta} = (\theta^1, \dots, \theta^n)$ over $H$ total update steps, linear parameter dependencies are captured by the empirical covariance matrix $\Sigma = \frac{1}{H-1}(\bm{\theta} - \bar{\bm{\theta}})^\top(\bm{\theta} - \bar{\bm{\theta}})$. 
To summarise stationary variability and coupling strength as a single scalar, we compute its Frobenius norm \citep{bottcher2008frobenius}:
\begin{equation}
    \|\Sigma\|_F = \sqrt{\sum_{i,j} \Sigma_{ij}^2}.
\end{equation}
Unlike trace-based metrics, the Frobenius norm incorporates cross-covariance terms, providing a compact descriptor of joint fluctuation complexity.

\subsection{(Maximum) Lyapunov Exponent Estimation}
To quantify local sensitivity to initial conditions, we estimate the maximal Lyapunov exponent $\lambda_{\max}$ from multivariate trajectories $\bm{\theta} = [\theta_h]_{h=1}^H \in \mathbb{R}^{H \times \Theta}$ via nearest-neighbour divergence. 
For each state $\theta_i$, its nearest neighbour $\theta_{j(i)}$ is found outside a Theiler window of $\pm w$ steps \citep{theiler1990estimating}. 
The Euclidean distance between forward trajectories at lag $z$, $d_i(z) = \| \theta_{i+z} - \theta_{j(i)+z} \|_2$, is tracked over $z \in [z_\text{min}, z_\text{max}]$, yielding:
\begin{equation}
    \lambda_{\max} \approx \frac{d}{dt} \langle \log d(z) \rangle.
\end{equation}
A positive $\lambda_{\max}$ denotes exponential divergence and chaotic learning dynamics.

\subsection{Recurrence and Correlation Plots}
Recurrence plots visualise phase-space revisitations by thresholding pairwise state distances:
\begin{equation}
    R_{ij} = \mathbf{1}\{ \|\theta_i - \theta_j\|_2 \le \varepsilon \},
\end{equation}
where $\varepsilon$ is calibrated to achieve a target recurrence rate (e.g., $8\%$). 
The binary matrix $R$ encodes temporal proximity: diagonal structures indicate epochs of predictability, while scattered points reflect stochastic or chaotic transitions \citep{eckmann1995recurrence, marwan2007recurrence, facchini2007curved}.

\subsection{Correlation Dimension (Fractal Dimension)}
We compute the correlation (fractal) dimension $D_2$ of the underlying attractor using the Grassberger–Procaccia algorithm \citep{grassberger1984dimensions}. 
Given a delay embedding $\Theta_h = [\theta_h, \theta_{h+\tau}, \ldots, \theta_{h+(m-1)\tau}] \in \mathbb{R}^m$ over $N$ reconstructed vectors, we evaluate the correlation sum across logarithmic radii $r$:
\begin{equation}
    C(r) = \frac{2}{N(N-1)} \sum_{i<j} \mathbf{1}\{ \|\Theta_i - \Theta_j\|_2 < r \}.
\end{equation}
In the linear scaling regime where $C(r) \propto r^{D_2}$, regression of $\log C(r)$ against $\log r$ yields $D_2$.

\subsection{Summary of Computed Metrics}
In summary, for each analysed trajectory the following quantities are reported:

\begin{itemize}
    \item Stationary distribution $\hat{p}(\bm \theta)$: empirical invariant measure.
    \item $\|\Sigma\|_F$: Frobenius norm of the covariance, capturing overall variability.
    \item $\lambda_\text{max}$: largest Lyapunov exponent, measuring chaotic divergence.
    \item Recurrence plots: graphical tools used to visualize the times at which a dynamical system returns to states similar to previous ones, as defined by a chosen error tolerance.
    \item $D_2$: correlation (fractal) dimension of the reconstructed attractor.    
\end{itemize}

Together, these diagnostics provide a comprehensive characterisation of the dynamical complexity, stationarity, and stability properties of the interacting reinforcement learning agents.

\subsubsection*{Acknowledgments}
\label{sec:ack}
James Rudd-Jones was supported by the Engineering and Physical Sciences Research Council through a Doctoral Training Partnership (DTP) PhD studentship (grant number EP/W524335/1 - Award 28684830).


\bibliography{main}
\bibliographystyle{rlj}


\end{document}